\documentclass[prd,twocolumn,showpacs,superscriptaddress]{revtex4}
\usepackage{epsfig}
\usepackage{graphicx}
\usepackage{amsmath,amssymb}

\newcommand{\comment}[1]{}

\begin{document}

\title{Reducing the weak lensing noise for the gravitational wave Hubble diagram using the non-Gaussianity of the magnification distribution}

\author{Christopher M. Hirata}
\email{chirata@tapir.caltech.edu}
\affiliation{Caltech M/C 350-17, Pasadena, CA 91125, USA}

\author{Daniel E. Holz}
\affiliation{Theoretical Division, MS-B227, Los Alamos National Laboratory, Los Alamos, NM 87545, USA}

\author{Curt Cutler}
\affiliation{Jet Propulsion Laboratory, M/S 169-327, 4800 Oak Grove Drive, Pasadena, CA 91109, USA}

\date{May 25, 2010}

\begin{abstract}
Gravitational wave sources are a promising cosmological standard candle because
their intrinsic luminosities are determined by fundamental physics (and are
insensitive to dust extinction).  They are, however, affected by weak lensing
magnification due to the gravitational lensing from structures along the line of
sight. This lensing is a source of uncertainty in the distance determination,
even in the limit of perfect standard candle measurements.  It is commonly
believed that the uncertainty in 
the distance to an ensemble of gravitational wave sources is limited by the
standard deviation of the lensing magnification distribution divided by the
square root of the number of sources.  Here we show that by exploiting the
non-Gaussian nature of the lensing magnification distribution, we can improve
this distance determination, typically by a factor of 2--3; we provide a fitting
formula for the effective distance accuracy as a function of redshift for
sources where the lensing noise dominates.
\end{abstract}


\maketitle

\section{Introduction}

In 1998 two groups measuring the luminosity distance-redshift relation (the
Hubble diagram) from Type Ia supernovae (SNe) reported that the expansion of the
Universe was accelerating
\cite{1998AJ....116.1009R, 
1999ApJ...517..565P}.  This discovery has stimulated a range of efforts to
measure the cosmic expansion history and assess whether it is consistent with a
cosmological constant or if alternatives such as quintessence are required.  The
Type Ia SNe continue to provide one of the most valuable constraints
\cite{2007ApJ...666..694W}, due to quality data at a range of redshifts, and the
lack of cosmic variance limitations that plague alternatives such as weak
gravitational lensing (WL) and baryon-acoustic oscillations (BAO) at low
redshift.

The advent of gravitational wave astronomy has prompted interest in gravitational wave (GW) sources as a standard candle.  Schutz \cite{1986Natur.323..310S} showed that a gravitational waveform from 
merging compact objects can be used to measure the distance to the source; a redshift obtained of an electromagnetic counterpart or host galaxy would then allow one to place the GW source on a luminosity 
distance-redshift relation.  The GW source method has the key advantage over other standard candles that its luminosity can be determined from fundamental physics, thus alleviating the common concern 
with standard candles that they could evolve with cosmic time.  They are also insensitive to dust opacity.  Finally, many proposed space-based gravitational wave detectors measure test mass separations 
directly in units of the laser wavelength, as opposed to supernovae, which measure relative distances and require an independent calibration ladder in 
order to measure absolute distances.  Thus GW sources offer the potential for a low-systematics
probe of the expansion history of the Universe.  Examples of such sources would include mergers of massive black holes, observable with the Laser 
Interferometer Space Antenna (LISA 
\footnote{{\tt http://lisa.jpl.nasa.gov/}}), and binary neutron star or stellar mass black hole 
binaries, observable with a more futuristic space-based detector in the $\sim 1$ Hz band such as the Big Bang Observer (BBO).  The latter, in particular, could potentially observe hundreds of thousands 
of neutron star-neutron star binary inspirals \cite{2009PhRvD..80j4009C}.

A final advantage is that the distance determination to a GW source is limited by the signal-to-noise of the measurement (and by partial degeneracies with binary 
parameters), as opposed to Type Ia SNe, which have a seemingly random $\sim 15$\% scatter in their luminosities even after the light curve stretch correction.  This means that the statistical power 
of GW 
sources may be interesting even if the number of usable sources is far less than the number of usable SNe.  In fact, for high signal-to-noise detections of GW sources, the 
distance determination should be limited not by the intrinsic dispersion of the source nor by the measurement error, but rather by weak lensing 
magnification: the intervening matter between us and the source will magnify the GW source and affect our measurement of the distance.  The apparent flux from the 
source is increased by some factor $\mu$, which is usually $\sim 1$, and the apparent distance $D_{\rm app}$ thus differs from the true distance $D$ according to 
$D_{\rm app}=D\mu^{-1/2}$.  This phenomenon of course occurs for all standard candles, and has long been recognized as an issue for SNe \cite{1997ApJ...475L..81W, 
1998PhRvD..58f3501H, 2005ApJ...631..678H}, but its importance relative to intrinsic dispersion is much greater for gravitational waves.  (Gravitational wave measurements with nearby sources or with lower 
signal-to-noise 
per source, such as the proposed binary progenitors of short gamma ray bursts \cite{2006PhRvD..74f3006D, 2009arXiv0904.1017N}, are much less affected since the lensing scatter is subdominant.)

The usual way of accounting for the magnification effect is to suppose that it adds in quadrature with the intrinsic-luminosity and apparent flux measurement contributions to the distance error.  That 
is,
\begin{equation}
\sigma_{\ln D^2} = \sqrt{\sigma_{\ln L}^2 + \sigma_{\ln F}^2 + \sigma_{\mu}^2},
\label{eq:basic}
\end{equation}
where $D$ is the distance, $L$ is the intrinsic luminosity, $F$ is the measured flux, and $\mu$ is the magnification; with $N$ sources, this uncertainty is reduced by a factor of $\sqrt N$ 
\cite{2005ApJ...629...15H, 2005ApJ...631..678H}.  Since the last term dominates for GW sources and is significant for high-$z$ SNe, there is great motivation to try to reduce it.  One possibility would 
be to try to construct an estimated magnification $\hat\mu$ from external data sets; the last term should then be replaced by $\sigma^2(\mu-\hat\mu)$.  Unfortunately, WL 
shear maps have too much high spatial frequency noise to be useful as a magnification estimator for point sources \cite{2003ApJ...585L..11D}, but galaxy maps are highly correlated with the mass 
distribution and may be able to reduce the lensing dispersion term by a factor of $\sim\sqrt3$ \cite{2006ApJ...640..417G, 2007ApJ...658...52J, 
2009A&A...493..331J}.  Maps of flexion (i.e. the gradient of the shear measured using the banana-
or trefoil-shaped distortion of a galaxy \cite{2005ApJ...619..741G}) could also be useful if very high source densities ($>100\,$arcmin$^{-2}$) can be
obtained \cite{2010MNRAS.tmp..245S}.

The purpose of this paper is to explore yet another method of reducing the lensing dispersion in Eq.~(\ref{eq:basic}).  Because the probability density function 
(PDF) of $\mu$ is highly non-Gaussian (technically non-$\Gamma$, as we show in Appendix~\ref{app:ineq}), our ability to centroid it using $N$ sources can be much 
better than $\sigma_{\mu}/\sqrt N$.  This is in fact a 
familiar result: 
as an extreme case, many distributions such as the Airy diffraction pattern in optical astronomy can be centroided even though their variances are formally infinite.  
In the case of lensing magnification, the variance of $\mu$ is often dominated by a long tail to positive values, corresponding to lines of sight that pass through 
a galaxy or its halo, whereas the information content is dominated by relatively empty lines of sight with $\mu-1$ sharply peaked around a slightly negative value.  
In such cases, the use of outlier-rejecting statistics not only removes sources with misidentified redshifts \cite{2009PhRvD..80j4009C} but also reduces the 
uncertainty in $D(z)$ for correctly identified sources.  (A similar point has been made in the recent paper by Shang \& Haiman \cite{2010arXiv1004.3562S}.)

This paper is organized as follows.  In Section~\ref{sec:centroid}, we discuss the information-theoretic limits on centroiding a distribution.  
Numerical results and simulations are presented in Section~\ref{sec:numerics}.
Section~\ref{sec:constraints} gives cosmological constraints obtainable with the reduced centroid errors
for BBO and for LISA.  
We conclude and briefly discuss systematics in Section~\ref{sec:discussion}.

We focus here on the problem of measuring $D(z)$ from GW sources.  However, the formalism is applicable to any standard candle, and we briefly discuss 
the implications for Type Ia supernovae.

\section{Centroiding a distribution}
\label{sec:centroid}

In this section, we consider the problem of determining the distance $D$ to a population of $N$ standard sources at some redshift $z$.  For simplicity, we consider 
first the case with no intrinsic dispersion in the source luminosity, and then generalize to the case with a known additional source of scatter (e.g. an intrinsic 
dispersion or measurement uncertainty).  For large $N$, the maximum likelihood estimator for $\ln D^2$ is asymptotically unbiased, and saturates the Cramer-Rao bound 
on the uncertainty given by the Fisher information.

\subsection{No intrinsic dispersion}

We consider a distribution of sources with some magnification probability $P(x)$, where $x=\ln\mu$.  The apparent distance to the source is
\begin{equation}
\ln D^2_{\rm app} = \ln D^2 - x.
\end{equation}
Our job is then straightforward: we are to estimate $\ln D^2$ from $N$ independent and identically distributed values of $\ln D^2_{{\rm app},i}$.  If $N$ is 
sufficiently large (how large will be investigated below), then we may use the Fisher information to determine how well we can measure $\ln D^2$.
For a single sample ($N=1$), the Fisher information is
\begin{eqnarray}
I_{\ln D^2}^{(1)} &=& \left\langle \left[ \frac d{d\ln D^2}\ln P(\ln D^2 - \ln D^2_{\rm app}) \right]^2\right\rangle
\nonumber \\
&=& \left\langle \left[ \frac d{dx}\ln P(x) \right]^2\right\rangle
\nonumber \\
&=& \int P(x) \left[ \frac d{dx}\ln P(x) \right]^2 dx.
\label{eq:I}
\end{eqnarray}
For multiple independent samples, the Fisher matrix simply adds so that $I_{\ln D^2}=NI_{\ln D^2}^{(1)}$.  For large $N$, the uncertainty in $\ln D^2$ would then be
\begin{equation}
\sigma(\ln \hat D^2) \approx \frac1{\sqrt{NI_{\ln D^2}^{(1)}}}.
\label{eq:sigmaoverhat}
\end{equation}

\subsection{Intrinsic dispersions and measurement uncertainties}

We now consider the case where the error $x$ in $\ln D^2_{\rm app}$ is determined not just by lensing, but also by an additional contribution such as intrinsic 
dispersion (for statistical standard candles such as supernovae) or flux measurement uncertainty (which exists for any standard candle).  We denote the lensing 
contribution by $x_1$ and the additional dispersion by $x_2$.  We assume these to be independent with probability distributions $P_1$ and $P_2$, so that the 
probability of $x$ is given by a convolution:
\begin{equation}
P(x) = \int P_1(x_1)P_2(x-x_1) dx_1.
\label{eq:Px}
\end{equation}
This assumption should be true for the case where $x_2$ is dominated by intrinsic dispersion, since the intrinsic luminosity scatter of a source physically cannot 
depend on the lens alignment.  It might be violated for the case of the measurement uncertainty since a magnified source will be detected at higher $S/N$ and thus is 
likely to have a smaller fractional error on the flux; however this is probably only significant for the strongly lensed sources, which do not dominate 
the 
information integral, Eq.~(\ref{eq:I}).

In this paper, the intrinsic dispersion/measurement uncertainty will be taken to be a lognormal distribution:
\begin{equation}
P(x_2) = \frac1{\sqrt{2\pi}\,\sigma_{x_2}} \exp\left[ -\frac{(x_2+\sigma_{x_2}^2/2)^2}{2\sigma^2_{x_2}} \right],
\end{equation}
with the offset $-\sigma_{x_2}^2/2$ designed to ensure $\langle e^{x_2}\rangle=1$.

The Fisher information for the convolved distribution, and for its improvement ratio, can be obtained from the usual formula, Eq.~(\ref{eq:I}).

\subsection{Estimators}

In the limit of large $N$, the maximum likelihood estimator for $\ln D^2$ achieves the Fisher information errors.  This estimator is given by the implicit equation
\begin{equation}
\sum_{i=1}^N w(\ln D^2_{{\rm app},i}-\ln\hat D^2) = 0,
\label{eq:MLE1}
\end{equation}
where
\begin{equation}
w(x) = -\frac d{dx}\ln P(x)
\label{eq:MLE2}
\end{equation}
is a weight function.

This can be compared to the ``conventional'' estimator based on flux-averaging \cite{2005ApJ...631..678H}, i.e. based on conservation of mean surface brightness 
$\langle\mu\rangle=1$, which implies $\langle D_{\rm app}^{-2}\rangle=D^{-2}$.  This approach gives another distance estimate,
\begin{equation}
\hat D_{\rm C}^{-2} = \frac1N\sum_{i=1}^N D_{{\rm app},i}^{-2},
\label{eq:C}
\end{equation}
where the subscript ``C'' is used to denote the conventional estimator.  Note that Eq.~(\ref{eq:C}) is model-independent in the sense that no functional form for the 
magnification PDF is assumed.  Since $D_{{\rm app},i}^{-2}=D^{-2}\mu$, the standard deviation of Eq.~(\ref{eq:C}) is $D^{-2}\sigma_\mu$, and the logarithmic 
uncertainty is
\begin{equation}
\sigma(\ln\hat D_{\rm C}^2) \approx \frac{\sigma_\mu}{\sqrt N}.
\label{eq:Cvar}
\end{equation}

The Fisher error, $[NI^{(1)}_{\ln D^2}]^{-1/2}$, is of course always less than or equal to Eq.~(\ref{eq:Cvar}).  We give an elementary proof of this inequality in 
Appendix~\ref{app:ineq}.  There we also show that equality holds only in the case where the magnification PDF is a $\Gamma$-distribution, 
Eq.~(\ref{eq:app-eq}).  We expect that in practice Eq.~(\ref{eq:sigmaoverhat}) should be a substantial improvement over Eq.~(\ref{eq:Cvar}) because
the $\Gamma$-distribution does not resemble a realistic magnification PDF, since it cuts off exponentially at large magnifications.  The $\Gamma$ 
distribution is also far more symmetric than ``real'' PDFs: it always has a normalized skewness
\begin{equation}
S'_3=\frac{\langle(\mu-1)^3\rangle}{\langle(\mu-1)^2\rangle^2}
\end{equation}
equal to $S'_3=2$.

\section{Numerical results}
\label{sec:numerics}

We have now completed our survey of the theory; it is now time to actually evaluate the Fisher information for realistic magnification PDFs.  We first describe the construction of
the lensing magnification PDFs and then display results.  Finally we simulate the effect of a finite number of sources on the maximum likelihood estimator.

\subsection{Lensing PDFs}

We use the stochastic universe method presented in Holz \&
Wald \cite{1998PhRvD..58f3501H} to calculate the lensing PDFs. This method
calculates the full (weak and strong) lensing distributions utilizing a Monte
Carlo code: the universe in the vicinity of a photon path is generated randomly, and
the lensing effects from the matter distribution are calculated
analytically. We approximate the matter in the universe as pure dark matter
smoothly distributed in NFW
halos \cite{1997ApJ...490..493N}, with the halo masses
drawn from the Sheth-Tormen mass function \cite{1999MNRAS.308..119S}, and with
cosmological parameters $\Omega_m=0.28$, $\Omega_\Lambda=0.72$, $h=0.7$, and
$\sigma_8=0.79$. The lensing distributions are relatively insensitive to
both the details of the lenses and the values of the cosmological
parameters \cite{2002ApJ...572L..15W}.

\begin{figure}
\includegraphics[angle=-90,width=3.2in]{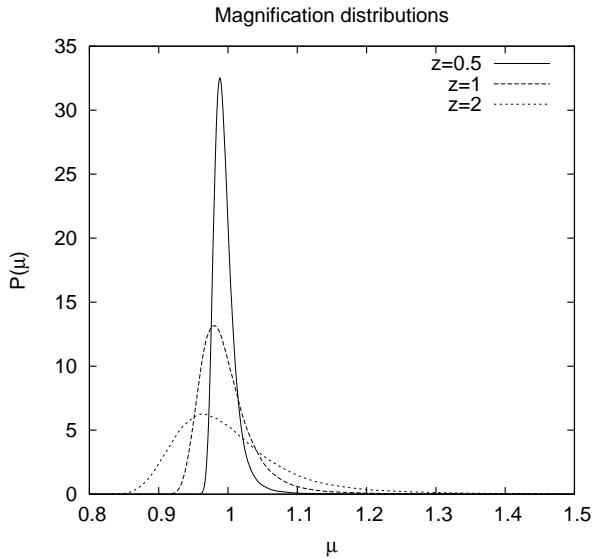}
\caption{\label{fig:magdist}The magnification distributions $P(\mu)$ at $z=0.5$, 1, and 2.}
\end{figure}

The Fisher analysis requires that the magnification PDF be smooth, since Monte Carlo noise results in a spurious, positive definite contribution to Eq.~(\ref{eq:I}).  We have
used several versions of the smoothing procedure.  The default procedure (used for most of the results in this paper unless otherwise specified) is to bin the lensing PDF in
bins of width $\Delta\mu=10^{-3}$.  Then a triangle-hat smoothing kernel is used, i.e.
\begin{equation}
P_{\rm smooth}(\mu) = \frac1{(S+1)^2}\sum_{j=-S}^S (S+1-|j|) P(\mu+j\Delta\mu).
\end{equation}
Since more smoothing is needed in the tail of the distribution than the peak, we generate two smoothed distributions $P_1$ and $P_2$ with different values of the smoothing $S_1$ and $S_2$.
The distributions are combined according to
\begin{equation}
P_{\rm smooth}(\mu) = \frac {P_2+cP_1}{1+c},
\end{equation}
where $c=e^{50(\mu-1.03)}$.  This results in an effective smoothing of $P_1$ at large $\mu$ and $P_2$ at small $\mu$ with smooth interpolation.  For $z=0.5$ we choose $(S_1,S_2)=(2,15)$; for $z=1$, 
(5,20); and for $z=2$, (10,30).
The Monte Carlo PDF is generated only out to $\mu=2$; above this, we assume a $P(\mu)\propto \mu^{-3}$ scaling as appropriate for large magnifications (near a caustic).  This matters little since this 
region contributes little to Eq.~(\ref{eq:I}).  The smoothed distributions at $z=0.5$, 1, and 2 are shown in Figure~\ref{fig:magdist}.

We have tried other methods of smoothing to ensure robustness.  For example, we have tried re-computing the Fisher integral, Eq.~(\ref{eq:I}), by 
fitting $P(\mu)$ with least-squares quadratic functions in intervals of width $\Delta\mu = 0.02$ (at $\mu>1+0.04z$) or 0.01 (at $\mu\le1+0.04z$), and 
using the fit to analytically compute $dP/dx$.  The integral is chopped off at the 0.1-percentile point of the distribution to avoid spurious effects 
(such as fits that pass through zero) since the quadratic polynomial is not a good approximation near the minimum value of $\mu$.  This procedure led 
to uncertainties that differed by at most 6\% from our fiducial smoothing procedure.

For completeness, we have also utilized a Savitzky-Golay smoothing filter, with
width $\Delta\mu=0.05$, and have found results differing by $\sim4\%$ from our
fiducial smoothing. 

\subsection{Fisher results}

\begin{figure}
\includegraphics[angle=-90,width=3.2in]{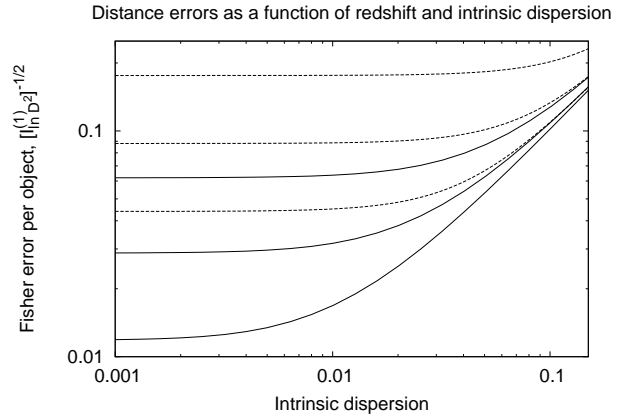}
\caption{\label{fig:centroid}The solid curves show the Fisher matrix error per source, $[I^{(1)}_{\ln D^2}]^{-1/2}$, as a function of intrinsic 
dispersion $\sigma_{x_2}$
for source redshifts of 0.5 (bottom), 1.0, and 2.0 (top).  The dashed curves show the error per source using the flux-averaging approach, 
Eq.~(\ref{eq:basic}).
}
\end{figure}

In Figure~\ref{fig:centroid}, we show the Fisher matrix error per source, $[I^{(1)}_{\ln D^2}]^{-1/2}$, as well as the flux-averaging error, 
Eq.~(\ref{eq:basic}).
For the flux-averaging error, we have used the approximation $\sigma_{\mu}\approx 0.088z$ \cite{2005ApJ...631..678H}.  The results are shown for 3 
redshifts, $z=0.5$, 1.0, and 2.0, and for a range of intrinsic dispersions $\sigma_{x_2}$.  For large intrinsic dispersion, lensing adds negligible 
additional dispersion and the error per source is $\sigma_{x_2}$.  For small $\sigma_{x_2}$, the lensing dispersion dominates, and we see that the 
Fisher matrix errors (solid curves) are factors of 2--3 below the flux-averaging errors (dashed lines).

%The improvement from using
%the maximum likelihood estimator (for large $N$) is shown in Figure~\ref{fig:improvement}.
%\begin{figure}
%\includegraphics[angle=-90,width=3.2in]{FisherImprovement}
%\caption{\label{fig:improvement}The improvement of the maximum likelihood approach over the flux-averaging approach, quantified by the ratio of standard deviations
%of the estimators for large $N$, i.e. we compute $(\sigma_\mu^2+\sigma_{x_2}^2)^{1/2}/[I^{(1)}_{\ln D^2}]^{-1/2}$.  The three curves show this improvement as a function of intrinsic dispersion 
%$\sigma_{x_2}$ for $z=0.5$, 1.0, and 2.0.}
%\end{figure}

In the case of the flux-averaging method, the variance per source is as noted above given by the quadrature sum,
\begin{equation}
\sigma^2_{\ln D^2} = \sigma^2_{\ln D^2}(\sigma_{x_2}=0) + \sigma_{x_2}^2.
\end{equation}
In the case of the centroiding method, no such exact relation holds.  However, it turns out that the quadrature sum approximation
\begin{equation}
[I^{(1)}_{\ln D^2}]^{-1} \approx [I^{(1)}_{\ln D^2}(\sigma_{x_2}=0)]^{-1} + \sigma_{x_2}^2
\label{eq:quadrature}
\end{equation}
has an error of at most 10\% over the range of redshifts $z\ge 0.5$ probed (the maximum error is at low redshift where the magnification PDF is most 
non-Gaussian).  By construction, Eq.~(\ref{eq:quadrature}) is exact when one or the other source of error dominates.

For an intrinsic scatter $\sigma_{x_2}\approx 0.15$ appropriate to Type Ia supernovae, the intrinsic scatter is dominant over the lensing even for the 
flux-averaging method for $z\le 1.7$.  At $z=1.7$ the centroiding method reduces the effective error per source $[I^{(1)}_{\ln D^2}]^{-1/2}$ from 0.22 
to 0.16, which is a modest improvement but not nearly what one finds for low dispersion gravitational wave sources.

\subsection{Finite sample size}
\label{ss:mle}

The maximum likelihood estimator is known to achieve the Fisher uncertainty only in the limit of large numbers of sources.
For a finite number of sources per redshift bin, Eq.~(\ref{eq:MLE1}) may be biased and may have a larger error than the Fisher information would suggest.
The {\em bias} in the estimator for $\ln D^2$ can be computed and subtracted using Monte Carlo simulations, assuming that $P(\mu)$ is known.
However, the uncertainty in $\ln D^2$ may be larger than the Fisher estimate.  Here we use Monte Carlo simulations to measure the scatter in $\ln\hat D^2$,
and show that for $N\ge 4$ sources the Fisher estimate is accurate to within $\le10$\%.

We have constructed our Monte Carlo simulations for any source redshift $z$ and intrinsic dispersion $\sigma_{x_2}$
by first drawing $N$ random deviates from the $P(\mu)$ distribution.  For each $\mu_i$, we obtain an estimated
luminosity distance $D_i^2 = D^2\mu_i^{-1}$.  Then we maximize
\begin{equation}
\ln L(\hat D) = \sum_{i=1}^N P(x=\ln\hat D^2-\ln D_i^2).
\end{equation}
Then $q=\ln\hat D^2-\ln D^2$ has a probability distribution that depends only on $N$ and $P(\mu)$ ($D$ trivially drops out).  A 95\% confidence interval on $\ln D^2$
can be obtained by taking $\ln\hat D^2$ and subtracting the 2.5th or 97.5th percentiles of the distribution of $q$, i.e. at 95\% confidence
\begin{equation}
\ln\hat D^2-q_{0.975}
<\ln D^2<
\ln\hat D^2-q_{0.025},
\end{equation}
where $q_\alpha$ is defined by $\int_{-\infty}^{q_\alpha} P(q)dq=\alpha$.
The width of the confidence interval is $q_{0.975}-q_{0.025}$.

In comparison, the usual assumption of Gaussian errors with width given by the Fisher calculation would suggest that the width of the 95\% confidence interval would be
$2\cdot1.960[NI^{(1)}_{\ln D^2}]^{-1/2}$, where 95\% of the probability in a unit normal distribution lies between $\pm1.960$.

\begin{figure}
\includegraphics[angle=-90,width=3.2in]{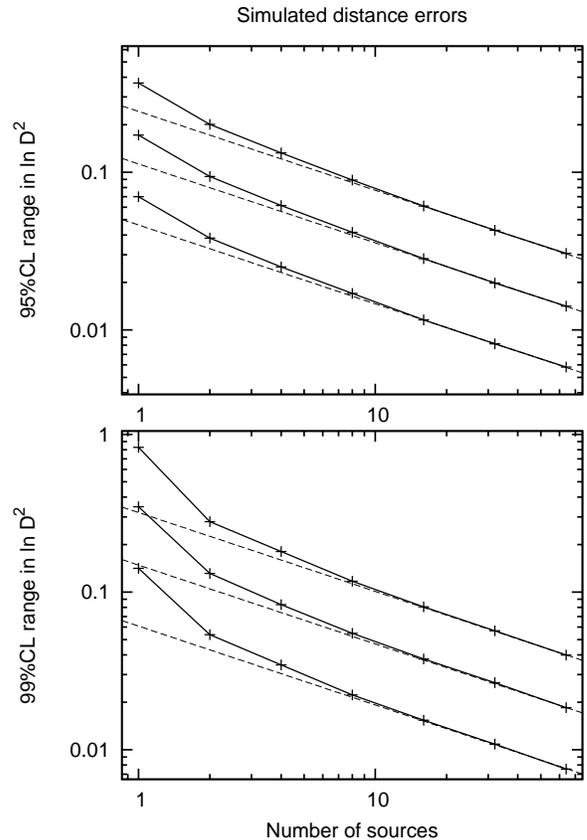}
\caption{\label{fig:nPlot}The full width of the 95\% and 99\% confidence ranges for $\ln D^2$
for source redshifts of 0.5 (bottom curve), 1.0, and 2.0 (top curve), as a function of the number of sources $N$.  Zero intrinsic dispersion is assumed.
The points are the results from Monte Carlo simulations.  The dashed lines are the Fisher predictions assuming Gaussian errors, i.e.
$2\cdot1.960[NI^{(1)}_{\ln D^2}]^{-1/2}$ and $2\cdot2.576[NI^{(1)}_{\ln D^2}]^{-1/2}$ respectively.  Note the good agreement of the Monte Carlo and Fisher
results for $N\ge4$.  This plot used $10^4$ simulations.
}
\end{figure}

The widths resulting from the full Monte Carlo procedure are compared with the Fisher calculation in Figure~\ref{fig:nPlot}.  As expected, the simulated errors are
larger than the Fisher prediction.  However the discrepancy drops rapidly for $N\ge4$.  This is because even 4 events are usually sufficient to identify and
reject the strongly magnified sources.

One would intuitively expect that the Fisher errors are approached more rapidly in the presence of intrinsic dispersion because this results in a $P(x)$ that
is more nearly Gaussian.  Indeed, this is what we find in simulations.  For example, at $z=1$ and $N=4$, we find that with no intrinsic dispersion the 99\% confidence
region is 1.12 times wider than the Fisher calculation suggests.  This factor drops to 1.09 if we impose an intrinsic dispersion of $\sigma_{x_2}=0.02$,
to 1.07 at $\sigma_{x_2}=0.05$, and 1.04 at $\sigma_{x_2}=0.1$.

\section{Cosmological constraints}
\label{sec:constraints}

We now consider the possible cosmological constraints from gravitational wave sources.  We begin by describing our methodology for computing parameter constraints (Sec.~\ref{ss:methods}).  We then turn 
to two specific examples: BBO (Sec.~\ref{ss:constraints-bbo}) and LISA (Sec.~\ref{ss:constraints-lisa}).

\subsection{Forecasting methodology}
\label{ss:methods}

It is straightforward to generalize Eq.~(\ref{eq:I}) to $N$ sources, at a range of range of redshifts $(z_1, z_2, \cdots, z_N)$,
and to a cosmological model depending on $N_P$ parameters.  We denote the cosmological parameters
% ($H_0$, $\Omega_m h^2$, etc.)
by $\{\lambda^{\alpha} \}_{\alpha = 1}^{ N_P}$.  Let $D_{{\rm app},i}$ represent the $N$ measured  luminosity-distances,
and let  $x_i \equiv  (\ln D(z_i)^2 - \ln D^2_{\rm app,i})$.
The magnifications for each source should be very close to statistically independent, since a gravitational wave detector sees the whole sky.  (Note that this is unlike the case of a supernova survey 
with an optical telescope, which inherently has a narrow field of view and hence depending on the survey strategy may suffer from correlated magnifications \cite{2006PhRvL..96b1301C}.)
Thus the $N_P$-dimensional Fisher matrix is
\begin{eqnarray}
I_{\alpha \beta} \!\!\! &=& \!\!\! \int P(x_1)P(x_2) \cdots P(x_N) \\ \nonumber
&& \!\!\! \Bigl[ \Big( \sum_{i=1}^N\frac{\partial x_i}{\partial \lambda^\alpha} \frac{d\ln P(x_i)}{dx_i} \Big)
\nonumber \\ && \!\!\! \times
 \Big( \sum_{j=1}^N\frac{\partial x_j}{\partial \lambda^\beta} \frac{ d\ln P(x_j) }{dx_j}\Big)\Bigr]
 dx_1 \cdots dx_N,
 \label{eq:IP2}
\end{eqnarray}
where we have used the fact that $P(x_1,x_2,\cdots,x_N) = P(x_1)P(x_2)\cdots P(x_N)$, since the $N$ measurements are
independent.  Using the fact that $\int P(x_i) dx_i = 1$ and hence that $\int P(x_i) [\partial \ln P(x_i)/\partial \lambda^{\alpha}] dx_i = 0$, it
is easy to see in the double-sum over $i$ and $j$, the terms with $i \ne j$ are all zero.  Hence our expression for $I_{\alpha \beta}$ reduces to
\begin{equation}
I_{\alpha \beta}= \sum_{i=1}^N \int P(x_i)\left[ \frac{d\ln P(x_i)}{dx_i} \right]^2 
\frac{\partial x_i}{\partial \lambda^\alpha}
\frac{\partial x_i}{\partial \lambda^\beta} \, dx_i.
\label{eq:IP}
\end{equation}
\noindent
Of course,  ${\partial x_i}/{\partial \lambda^{\alpha}}$ is just  ${\partial ({\rm ln} D^2_i)}/{\partial \lambda^{\alpha}}$.
Thus we arrive at our final expression for the Fisher matrix:
\begin{equation}
I_{\alpha \beta}= \sum_{i=1}^N  I^{(1)}_{{\rm ln} D^2}( z_i)\Big[\frac{\partial (\ln D^2(z_i))}{\partial \lambda^\alpha} \frac{\partial (\ln D^2(z_i))}{\partial \lambda^{\beta}}\Big].
\label{eq:IP3}
\end{equation}
The information for a single source $I^{(1)}_{{\rm ln} D^2}( z_i)$ is simply the integral, Eq.~(\ref{eq:I}), where the probability distribution for $x$ contains lensing noise and (if significant) 
measurement noise as well.

\begin{figure}
\includegraphics[angle=-90,width=3.2in]{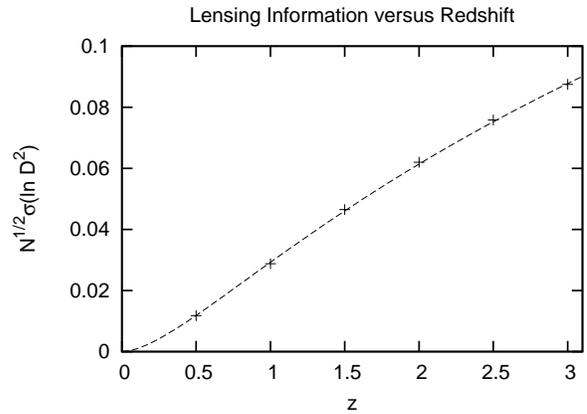}
\caption{\label{fig:fitplot}The uncertainty per source on the Hubble plot, $[I^{(1)}_{{\rm ln} D^2}( z)]^{-1/2}$, as a function of redshift.
The points are obtained from evaluation of the information integral over $P(\mu)$, while the curve is the fit of Eq.~(\ref{eq:mufit}).
}
\end{figure}

To rapidly estimate $I^{(1)}_{{\rm ln} D^2}( z)$ for any $z$ (up to $z =3$),
we  (i) calculated the magnification distribution $P_z(\mu)$, for  redshifts $z\in\{0.5, 1, 1.5, 2, 2.5, 3\}$ using the method of 
Holz \& Wald~\cite{1998PhRvD..58f3501H}, (ii) used (smoothed versions of) these distributions to calculate
$I^{(1)}_{{\rm ln} D^2}(z)$ for these $6$ redshifts, using Eq.~(\ref{eq:I}), and then (iii) fit these results to a smooth curve that has the correct asymptotics (going to $0$ as $z\rightarrow 0$ and 
going to a constant as $z\rightarrow \infty$).
We find the following to be a suitable fit:
\begin{equation}
\label{eq:mufit}
%\Big[I^{(1)}_{{\rm ln} D^2}( z)\Big]^{-1/2} = -0.0073 + 0.0396\, z - 0.00284 z^2
\Big[I^{(1)}_{{\rm ln} D^2}( z)\Big]^{-1/2} = C
  \left[ \frac{1-(1+z)^{-\beta}}\beta \right]^\alpha,
\end{equation}
\noindent
where $C = 0.066$, $\beta = 0.25$, and $\alpha = 1.8$.  This function is shown in Figure~\ref{fig:fitplot}.  Note that we have not explored its validity at $z>3$.

At low redshifts, the peculiar velocity 
error dominates; assuming a width of 300 km$\,$s$^{-1}$ (e.g. \cite{2006A&A...447...31A}) gives an additional contribution to $\sigma_{x_2}$ of twice 300 
km$\,$s$^{-1}$ divided by the 
Hubble velocity $cz$, which is $0.002z^{-1}$.  We have approximated this error by a quadrature-sum with the lensing + measurement noise, which our tests of 
Eq.~(\ref{eq:quadrature}) suggest will not be in serious error.

In some cases, a very large number of sources will be observed (possibly of order $10^5$ for BBO).  In such cases, it is appropriate to bin the sources into redshift slices as is often done for supernova 
forecasts \cite{2009arXiv0901.0721A}.  Given a redshift distribution $\Pi(z)$,
we bin sources into redshift slices of width $\Delta z=0.1$, and the number of sources in each bin is computed according to $N_i = N_{\rm src} \Pi(z)\Delta z$.  We have cut off the bins at redshifts 
below $z_{\rm min}$, where
$z_{\rm min}$ is defined such that we expect 1 source at $z<z_{\rm min}$, i.e. $\int_0^{z_{\rm min}} \Pi (z) dz = N_{\rm src}^{-1}$.  This is to prevent a mission that
collects a small number of low-$z$ sources from taking advantage of a highly precise ``constraint'' obtained locally ($z\ll 1$) from e.g. 0.01 sources
with ultra-precise distances.

Our parameter space $\{\lambda^\alpha\}$ includes the present-day densities of baryons $\Omega_{\rm b}h^2$, matter $\Omega_{\rm m}h^2$, and dark energy 
$\Omega_{\rm de}h^2$, as well as the spatial curvature $\Omega_{\rm K}h^2$.  The equation of state (pressure:density ratio) of the dark energy is 
described by the 2-parameter model
\begin{equation}
w_{\rm de}(a) \equiv \frac{P_{\rm de}(a)}{\rho_{\rm de}(a)} = w_0 + w_a(1-a),
\end{equation}
where the parameters are $(w_0,w_a)$ and the fiducial ``cosmological constant'' model has $w_0=-1$ and $w_a=0$.  We also include the primordial 
spectrum of Gaussian adiabatic cosmological perturbations, assumed to be a power law (2 parameters: amplitude and spectral index), which are required 
when combining GW data with the CMB or weak lensing; and the absolute magnitude of a Type Ia supernova, required when including the supernova Hubble 
diagram.  This gives a 9-dimensional parameter space.  Note that the Hubble constant $H_0$ is not an additional parameter since it is determined by 
$\Omega_{\rm m}h^2$, $\Omega_{\rm de}h^2$, and $\Omega_{\rm K}h^2$.

We run our forecasts both internal to the GW method, and in combination with other methods of probing cosmology; the latter cases include the ``Stage 
III'' (i.e. near-term ground based) results for 
supernovae, weak lensing, and baryon-acoustic oscillations, and the
{\slshape Planck} CMB constraints, as forecast by the Figure of Merit Science Working Group (FoMSWG) 
\cite{2009arXiv0901.0721A}.

We compute the Figure of Merit (FoM) defined by the Dark Energy Task Force (DETF) \cite{2006astro.ph..9591A} for several combinations of future 
data sets.  This FoM is defined to be proportional to the inverse-area of the error ellipse in the $(w_0,w_a)$ plane, i.e.
\begin{equation}
{\rm FoM}_{\rm DETF} \equiv \frac1{\sigma(w_0)\sigma(w_a)\sqrt{1-\rho^2(w_0,w_a)}},
\end{equation}
where $\rho(w_0,w_a)$ is the correlation coefficient.  The DETF Figure of Merit is not a unique (or even necessarily the best) way to present the 
performance of a dark energy experiment -- see the discussion in the FoMSWG report \cite{2009arXiv0901.0721A} -- but it is
well suited to our objective here, which is to show that our magnification PDF centroiding method leads to significantly tighter dark energy 
constraints from GW sources.

\subsection{Example: BBO}
\label{ss:constraints-bbo}

We consider a population of sources with the redshift distribution of 
Ref.~\cite{2009PhRvD..80j4009C}, appropriate to neutron star-neutron star (NS-NS) binaries.  Two cases are considered for the error of the distance 
determination: an ideal case (IDEAL, $\sigma_{x_2}=0$), and an error of 1.4$z$\% (NSNS, $\sigma_{x_2}=0.028z$), with the latter appropriate to BBO parameter forecasts 
\cite{2009PhRvD..80j4009C}.  For each of these cases, we consider two subcases for the distance determination, the flux-averaging method (AVE) and the centroiding 
method described here (CEN).
For the centroiding case, we used the method described above, while
for the flux-averaging cases, the $\ln D^2$ uncertainties per 
source are $0.088z$ (IDEAL.AVE) or $\sqrt{0.088^2+0.028^2}\,z$ (NSNS.AVE).

\begin{figure}
\includegraphics[angle=-90,width=3.2in]{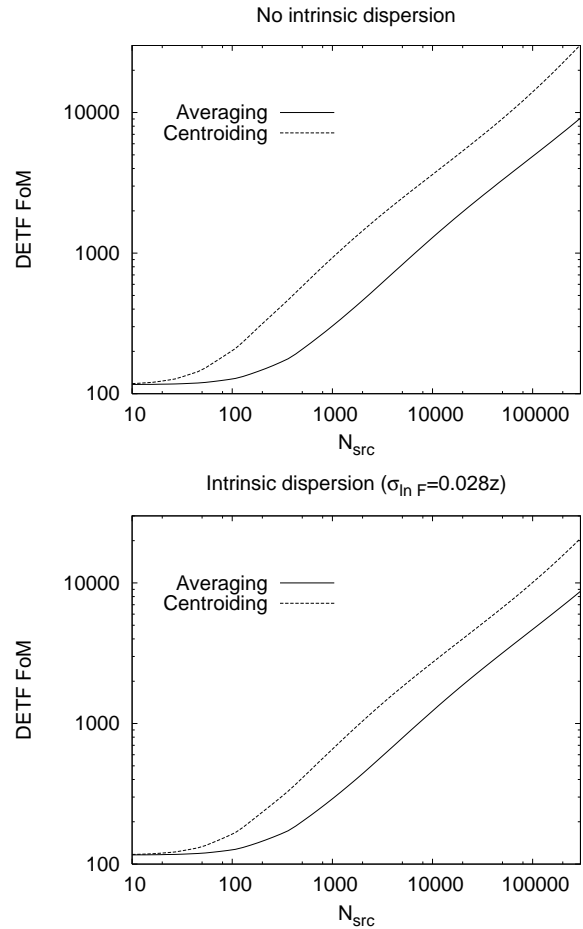}
\caption{\label{fig:fom3}The DETF FoM as a function of the number of gravitational wave sources $N_{\rm src}$ used.  We also include both the {\slshape Planck} mission and next-generation ground-based 
dark energy projects (Stage III).
The highest $N_{\rm src}$ value plotted corresponds to $N_{\rm src}=3\times 10^5$, the rough number expected for BBO.}
\end{figure}

In Figure~\ref{fig:fom3}, we show the DETF Figure of Merit for the combined constraints.  We have {\em not} included any systematics in 
the GW constraints.  This is of course an optimistic case, and it is not yet clear whether the systematic errors can be made negligible for a BBO-class 
mission.  For example, while the physics of the GW source (the 2-body problem in general relativity) is ``clean,'' there are possible systematic errors 
associated with (i) the theoretical predictions for the magnification PDF $P(\mu)$, particulary
associated with baryonic physics on small scales; and (ii) the strain calibration of a BBO-type detector \cite{2009PhRvD..80j4009C}.

We can see that an improvement of a factor of $\sim 2$ in the FoM is possible
with $\sim 130$ well-measured sources, and an order of magnitude improvement (comparable to that promised by various Stage IV projects such as the more grandiose versions of
JDEM \footnote{{\tt http://jdem.gsfc.nasa.gov/}}) is possible with $\sim 1500$ sources.  We also see that using the flux averaging rather than the centroiding
results in a factor of $\sim 5$ increase in the number of sources required to reach a given DETF FoM for the IDEAL case (and a factor of $\sim 3$ for 
the NS-NS case).

%We also consider what happens if gravitational wave sources are used by themselves.  By the time BBO (or even LISA) flies, there will have been many ground-based dark 
%energy projects, but one may assess the robustness of an overall dark energy program in part by examining how well one can do with each dark energy technique 
%\cite{2006astro.ph..9591A}.  Thus, in Figure~\ref{fig:fomred}, we show the DETF FoM for gravitational wave sources only and in combination with {\slshape Planck}.
%A BBO-type mission ($3\times 10^5$ sources; NSNS.CEN) plus {\slshape Planck} yields a theoretical FoM of $1.3\times 10^4$, far greater than any of the Stage 
%III or Stage IV dark energy projects \cite{2006astro.ph..9591A}.  Nevertheless, we can see that the combination with {\slshape Planck} is essential, as the difference 
%between the with-{\slshape Planck} and without-{\slshape Planck} FoMs is at least an order of magnitude.

%\begin{figure}
%\includegraphics[angle=-90,width=3.2in]{fomred}
%\caption{\label{fig:fomred}The DETF FoM as a function of the number of gravitational wave sources $N_{\rm src}$ used.  The top panel is the constraint for gravitational waves only, while the bottom 
%panel 
%includes {\slshape Planck}.  No other low-redshift dark energy probes (supernovae, weak lensing, or baryon oscillations) are assumed.}
%\end{figure}

\subsection{Example: LISA}
\label{ss:constraints-lisa}

As a second application of the main ideas in this paper, we consider 
how well cosmological parameters might be constrained by LISA observations 
of coalescing massive black hole binaries (MBHBs).  This question has been 
considered by several authors, including~\cite{2005ApJ...629...15H,2006ApJ...637...27K,2008PhRvD..77d3512M,2008ApJ...677.1184L,2010arXiv1001.3099V}, 
with the importance of weak lensing as the dominant effect 
in limiting LISA's distance-measurement accuracy first being stressed by 
Hughes \& Holz~\cite{2003CQGra..20S..65H}. The main result of this paper -- that 
previous analyses considerably underestimated the improvement in 
$D_L$-accuracy that comes from combining several measurements -- suggests a re-examination of the LISA case.

%for $N \agt 4 $ lensed sources at the same
%redshift, the error in $D_L$ is actually considerably smaller than  $N^{-1/2}
%
%improved accuracy available from "centroiding the magnification
%distribution", 

To provide some context: LISA will be capable of detecting merging 
MBHBs with masses in the range $\sim 10^3 - 10^6 M_\odot $ out to $z \sim 20$.  
Estimates of MBHB merger rates vary by several orders of magnitude, depending mostly on the fraction of dark-matter halos 
that have MBHs in their cores at redshifts $z \sim 10-20$.
For example, merger-tree models due to Volonteri predict that LISA should detect $\sim 30$ MBHBs/yr, mostly at high redshift ($z>4$)
\cite{2009CQGra..26i4027A}.
LISA's angular resolution will typically
be a few degrees or worse, so to uniquely identify the host galaxy will generally require some corresponding electromagnetic 
outburst.  Several possible mechanisms for generating outbursts have been explored, including (i) excitation of a shared accretion
disk due to the rapid mass loss and/or velocity kick when the binary merges (a consequence of the energy and momentum carried off in GWs)
\cite{2008ApJ...676L...5L, 2009AAS...21344905B, 
2009arXiv0910.0014C}, (ii) a steep rise
in the accretion rate starting months to years after the merger
\cite{2005ApJ...622L..93M}, and (iii) a pre-merger burst due to shepherding of the disk around one of the progenitor black holes 
\cite{2009arXiv0906.0825C}.  But for accurate knowledge
of the intensity, spectrum, and time-profile of such electromagnetic outbursts, we may well have to wait until LISA flies.  Lacking 
robust predictions regarding electromagnetic outbursts, the LISA community has tentatively adopted the criterion that an MBHB merger is
promising for precise localization if the LISA $1\sigma$ error ellipse on the sky is $\lesssim 10$ deg$^2$,
which is roughly the field of view of the
planned Large Synoptic Survey Telescope (LSST) \footnote{{\tt http://www.lsst.org/lsst}}.   Applying this criterion to Volonteri's population 
models, for example, one finds that 
$\sim \frac13$ of LISA's detected MBHBs could be positioned to within $\lesssim 10$ sq. degrees.   Even if only $\sim 20\%$ of these ``promising''
events could actually have their redshift determined, this would still lead to of order $10$ points on the $D_L-z$ curve where the luminosity distances follow from
fundamental physics (the 2-body problem in GR).  Errors in  $D_L$ due to noise (both instrumental noise and the confusion background from $\sim 3 \times 10^8$ compact  galactic
binaries) will typically be only $\sim 1-2\%$, even before incorporating the extremely accurate sky-localization provided by an EM counterpart.  Using the precise EM localization will
typically reduce this uncertainty by a factor $\sim 2$~\cite{2010arXiv1001.3099V}. Therefore we are in a regime where WL magnifications strongly 
dominate the $D_L$ errors. 

Given the large uncertainties, in this paper we adopt a very simple
model for the $z$-distribution of localizable sources, which is as follows.  We take
the rate of MBHB mergers (per unit comoving volume, per unit proper time) to be some constant $\dot n$, with the universe evolving
according to a flat $\Lambda$CDM model, with our fiducial values $(\Omega_{\Lambda} = 0.744, \Omega_m = 0.256)$.
Then the redshift distribution of the of the binary sources whose GWs are arriving at LISA (over an observation time $T_{\rm obs}$) is:
\begin{equation}
\frac{dN}{dz} = 4\pi \dot n T_{\rm obs} \frac{r^2(z)}{(1+z)H(z)},
\label{eq:dNdz}
\end{equation} 
where $r(z)$ is the comoving distance to redshift $z$ and $H(z)$ is the Hubble rate.

%This distribution is shown in Fig.~\ref{rate}, normalized to $\dot n = H^{-4}_0$ and $T_{\rm obs} = 5\,$yr.
%[[CH: Is this the normalization? Should check but we re-normalize it later.]]
%\begin{figure}
%\includegraphics[angle=0,width=3.2in]{dndz}
%\caption{\label{fig:rate} The z-distrubution of MBH mergers detected by LISA, assuming the rate (per unit co-moving volume, per unit proper time) is constant. Here we have normalized to $\dot n = 
%H^{-4}_0$ and $T_{obs} = 5$yr.  [[CC:Guys, this is just a sample figure;
%if we decide to keep it, I can make a prettier version.]]}
%\end{figure}

We restrict attention to mergers
at $z<z_{\rm max} = 3$, and we assume that some constant fraction $F$ of all mergers in this redshift range
can be associated with an EM outburst that identifies the host galaxy.  The rough justification for limiting our attention to redshifts $z < z_{\rm 
max}$ is that as $z$ increases it clearly becomes 
harder to identify a counterpart, both
because the GW SNR is lower (which increases the size of the error box) and because any EM outburst is fainter (and at $z\gtrsim 7.5$ is 
completely 
obscured in the LSST bandpasses by intergalactic 
Lyman-$\alpha$ absorption); we crudely model this fall-off by a Heaviside function.  We generate parameter constraints by the method described in Section~\ref{ss:methods}.

\begin{figure*}
\includegraphics[angle=-90,width=6.5in]{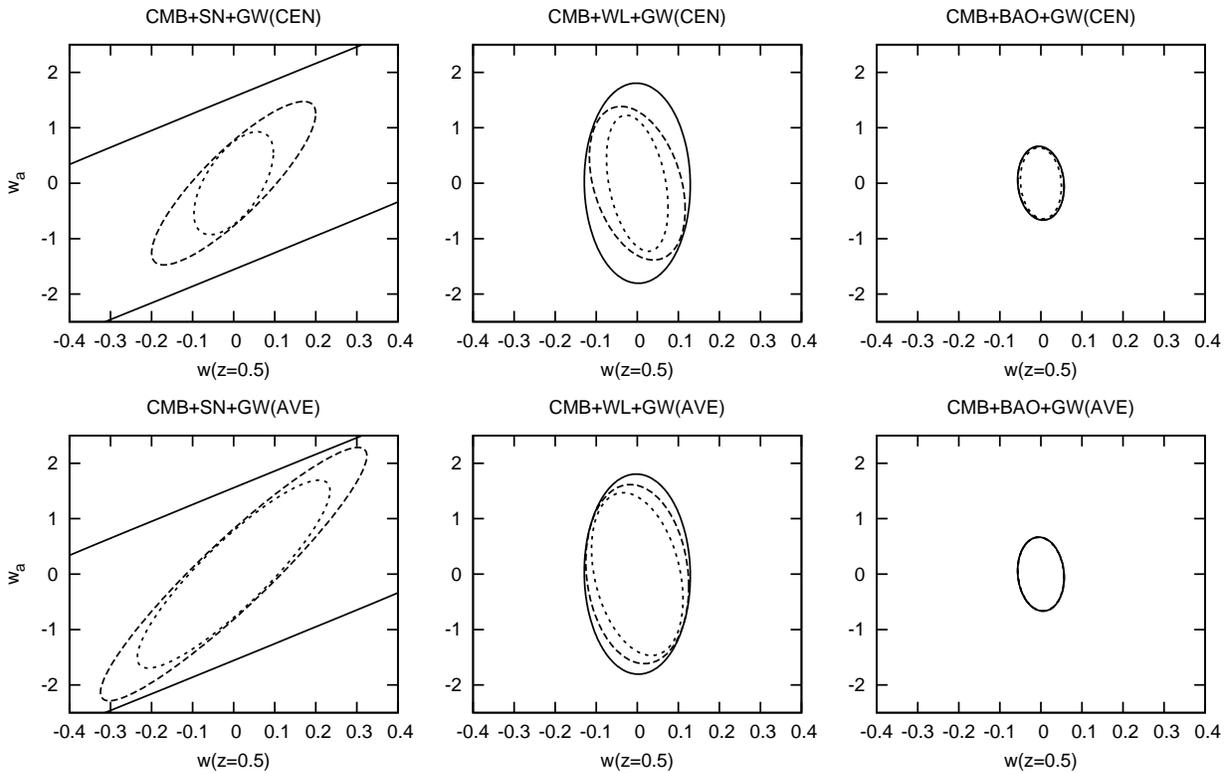}
\caption{\label{fig:wplot}The constraints on the $(w_0,w_a)$ model.  The solid ellipses are the forecast 68\%\ confidence contours ($\Delta\chi^2 = 2.28$) for {\slshape Planck} plus the indicated 
cosmological probe (SN, WL, or BAO) at Stage III using the
FoMSWG Fisher matrices \cite{2009arXiv0901.0721A}.  The dashed and dotted ellipses incorporate LISA constraints assuming either 10 usable sources at $z<3$ (dashed) or 30 sources (dotted).  The upper 
panels shows results using the centroiding technique (CEN), while the bottom panels use flux averaging (AVE).  The horizontal axis is plotted as $w(z=0.5)=w_0+\frac13w_a$ instead of $w_0$ in order to 
avoid long diagonal contours; note that this transformation preserves the areas of the ellipses.}
\end{figure*}

The resulting parameter constraints are shown in Figure~\ref{fig:wplot} for both the case of 10 and 30 usable electromagnetic counterparts.  The addition of the GW constraint does not significantly 
improve the Stage III DETF Figure of Merit -- for 30 sources it raises it from 116 to 130 (although we note that the investigation of GW dark energy constraints is still in its early days and further 
improvements may be possible).  However,
one may assess the robustness of an overall dark energy program in part by examining how well one can do with each dark energy technique
\cite{2006astro.ph..9591A}.  We therefore show in Figure~\ref{fig:wplot} constraints for the CMB+SN+GW, CMB+WL+GW, and CMB+BAO+GW cases.  The gravitational wave constraints make large improvements when 
combined with the supernovae (they partially break the $w_a-\Omega_{\rm K}$ degeneracy by extending the Hubble diagram to higher redshifts) and weak lensing.  Less improvement is seen with the BAO model 
because the BAO already provide some distance constraints in the $z>1$ range.  As one can see by comparing the top and bottom rows of the figure, the parameter constraint improvements are much more 
impressive when using the centroiding than the flux-averaging method.

\section{Discussion}
\label{sec:discussion}

The luminosity distance-redshift relation is one of the oldest and most fundamental probes of cosmology, and future gravitational wave detectors offer the possibility of measuring it accurately using 
binary inspirals whose luminosity can be calculated directly from measured parameters and fundamental physics.  These sources are however affected by weak gravitational lensing by intervening 
inhomogeneities in the cosmic mass distribution.  This introduces changes of typically a few percent (but occasionally much larger) in the flux, while not significantly affecting the redshift, and thus 
provides a source of noise in the $D(z)$ relation.  We have shown in this paper that exploiting the full power of the likelihood function can reduce this noise: the noise in the $D(z)$ relation is not 
limited by the lensing dispersion divided by the square root of the number of sources, but rather can be less by a factor of 2--3 if one centroids the distribution of apparent distances using the known 
non-Gaussian form of the lensing magnification PDF.

We have not discussed here the systematic errors associated with using large numbers of gravitational wave sources for precision low-redshift cosmology, as suggested for BBO.  While the signal itself is 
expected to be clean, there are 
potential sources of systematic error.  Some of these, such as strain calibration, coherent peculiar velocities at low redshift \cite{2006PhRvD..73j3002C, 2006PhRvD..73l3526H}, and host redshift 
misidentification, exist irrespective of the method used to estimate the true $D(z)$ from a collection of weakly lensed GW sources with their apparent fluxes and redshifts.  However, the issue of 
uncertainties in the magnification PDF is worth discussing here.  It may seem at first glance that the flux-averaging technique is more robust than the centroiding technique described here, because it 
relies only on the flux conservation constraint $\int_0^\infty P(\mu)d\mu=1$.  This may not be the case for three reasons.  One is that in order to remove misidentified host galaxies, it 
is likely that a BBO-type mission would reject outliers from the $D(z)$ relation \cite{2009PhRvD..80j4009C}.  This outlier rejection would eliminate the the tails of the 
magnification distribution with $|\ln\mu|>(0.4\ln10)\Delta M$, where $\Delta M$ is the half-width of the cut in magnitudes.  Since $P(\mu)$ is highly asymmetric, with the large-$\mu$ tail much stronger 
than the small-$\mu$ tail, it follows that outlier rejection will result in the conditional probability $\int_0^\infty P(\mu|{\rm accept})d\mu<1$ and hence give a positive bias in $D(\mu)$.  This can be 
corrected, but it requires knowledge of the contribution to $\int_0^\infty P(\mu)d\mu$ from the strong-magnification tail.  A second 
reason has to do with strong lensing: flux conservation implies that the magnification satisfying $\int_0^\infty P(\mu)d\mu=1$ is the {\em total} magnification of all of the images.  However, since 
strong-lens time delays are often measured in months (and even longer if the strongly de-magnified central image is significant), it is likely that for many BBO sources there will be additional images 
whose time delay places them outside the BBO mission lifetime.  Finally, in obtaining a successful host redshift (or identifying a source with the correct host rather than another object nearby on the 
sky), there is likely to be a bias in favor of brighter sources, which results in a success probability that has some dependence on the magnification, $P({\rm success})\sim\mu^\beta$.  The presence of 
lensing dispersion then results in a bias in the mean flux of $\sim\beta{\rm Var}\,\mu$; since Var$\,\mu$ is of order $10^{-2}$ at $z\sim 1$, such biases will likely be far above the BBO statistical 
errors, and will have to be corrected using knowledge of $P(\mu)$.  Therefore, even the flux-averaging method is sensitive to the particular distribution $P(\mu)$.  The problem of how well $P(\mu)$ can 
be determined via theory (particularly in the presence of baryonic effects), or reduced to a parametric model whose parameters can be simultaneously fit using BBO, is left to future work.

In this paper, we have not attempted to reduce the lensing dispersion by using external information, i.e. only the shape of $P(\mu)$ was assumed.  The main external 
sources of information that are commonly considered are weak lensing shear maps and galaxy maps.  In principle the way one would incorporate this information would be 
to write a conditional probability density, e.g. $P(\mu|g)$, where $g$ represents the galaxy density field.  Galaxy maps have been shown to be helpful because they 
contain information about the small scales that dominate the lensing variance \cite{2006ApJ...640..417G, 2007ApJ...658...52J, 2009A&A...493..331J}; however the 
conditional probability distribution $P(\mu|g)$ may be very hard to determine theoretically at the required precision.  For example, despite recent advances in 
determining the relation between galaxy luminosity and host halo mass (a key quantity of interest if one is trying to infer the matter density field from galaxy 
observations) using clustering and lensing data \cite{2005ApJ...630....1Z, 2006MNRAS.368..715M}, a measurement of the scatter in this relation is not yet possible, 
and the full distribution of this scatter -- required if one is going to compute $P(\mu|g)$ -- is woefully underconstrained.  Nevertheless, for a mission such as LISA 
that may have only a limited number of usable sources and hence may be dominated by statistical errors due to weak lensing, this may be a useful approach.  The weak 
lensing field is sensitive only to the matter distribution, and so one could imagine that it would be profitable to utilize the smoothed convergence field, 
$\kappa_{\rm sm}$, and attempt to centroid the conditional density $P(\mu|\hat\kappa_{\rm sm})$. WL has traditionally been viewed as not useful for de-lensing of GW 
sources because most of the lensing variance comes from small scales where weak lensing measurements are noisy \cite{2003ApJ...585L..11D}.  This conclusion should be 
revisited in future work using the centroiding technique; in particular, this small-scale structure contributes strongly to the high-magnification tail of 
$P(\mu|\hat\kappa_{\rm sm})$, and it is not yet known what happens to the Fisher information, which depends largely on the width of the peak of $P(\mu|\hat\kappa_{\rm 
sm})$.

\section*{Acknowledgements}

We acknowledge useful conversations with \'Eanna Flanagan and Samaya Nissanke.

C.H. is supported by the U.S. Department of Energy (DE-FG03-92-ER40701), the National Science Foundation (AST-0807337), and the Alfred P. Sloan 
Foundation.
D.H. acknowledges support from the LDRD program at LANL.
C.C.'s work was carried out at the Jet Propulsion Laboratory,
California Institute of Technology, under contract to the
National Aeronautics and Space Administration.  He acknowledges
support from a JPL Research and Technology Development
grant, as well as support from NASA grant NNX07AM80G.

Copyright 2010.  All rights reserved.

\appendix

\section{Information improvement}
\label{app:ineq}

This appendix is dedicated to proving the information-bounding inequality:
\begin{equation}
[I_{\ln D^2}^{(1)}]^{-1/2} \le \sigma_\mu
\label{eq:infobound}
\end{equation}
for any distribution such that $\langle\mu\rangle=1$.  We also show that equality holds only for the $\Gamma$-distribution:
\begin{equation}
P(\mu) = \frac{\alpha^\alpha}{\Gamma(\alpha)}\mu^{\alpha-1}e^{-\alpha\mu}.
\label{eq:app-eq}
\end{equation}

We begin by considering the functions
\begin{equation}
f(x) = (e^x-1)\sqrt{P(x)}
\label{eq:f}
\end{equation}
and
\begin{equation}
g(x) = -2\frac d{dx}\sqrt{P(x)}.
\label{eq:g}
\end{equation}
We now use the Cauchy-Schwarz inequality, $[\int f(x)g(x) dx]^2\le [\int f^2(x)dx][\int g^2(x)dx]$.  It is readily apparent that $\sigma_\mu^2 = \int f^2(x)dx$ since 
$\mu=e^x$ and $\langle\mu\rangle=1$.  We 
can also see that
\begin{eqnarray}
\int g^2(x)dx &=& \int \left[2\frac d{dx}\sqrt{P(x)}\right]^2 dx
\nonumber \\
&=& \int \left[ \sqrt{P(x)} \frac d{dx}\ln P(x) \right]^2 dx
\nonumber \\
&=& I_{\ln D^2}^{(1)}.
\end{eqnarray}
Finally,
\begin{eqnarray}
\int f(x)g(x) dx &=& -2\int (e^x-1)\sqrt{P(x)}\frac d{dx}\sqrt{P(x)} dx
\nonumber \\
&=& -\int (e^x-1)\frac d{dx}P(x) dx
\nonumber \\
&=& \int e^x P(x) dx = \langle\mu\rangle=1,
\end{eqnarray}
where in the third equality we have used integration by parts and noted that the surface terms vanish since in order to be normalized the probability distribution 
must vanish faster than $x^{-1}$ as $x\rightarrow\pm\infty$.  The Cauchy-Schwarz inequality then states
\begin{equation}
1\le \sigma_\mu^2I_{\ln D^2}^{(1)},
\end{equation}
thereby proving Eq.~(\ref{eq:infobound}).

Equality holds if and only if $g(x)=\alpha f(x)$ for some constant $\alpha$.  Examining Eqs.~(\ref{eq:f}) and (\ref{eq:g}) shows that equality is 
thus equivalent to a first-order ordinary differential equation for $\sqrt{P(x)}$,
\begin{equation}
-2\frac d{dx}\sqrt{P(x)} = \alpha(e^x-1)\sqrt{P(x)},
\end{equation}

~\\ \noindent which has the solution
\begin{equation}
P(x) \propto \exp[-\alpha(e^x-x)].
\end{equation}
Using $P(\mu) = P(x) dx/d\mu = P(x)/\mu$, we may write this as a function of $\mu$:
\begin{equation}
P(\mu) \propto \mu^{\alpha-1}e^{-\alpha\mu}.
\end{equation}
This equation is easily normalized and is given by Eq.~(\ref{eq:app-eq}).  By inspection its first moment is indeed $\langle\mu\rangle=1$, and its variance is 
Var$\,\mu=\alpha^{-1}$.

We note that for small variance (large $\alpha$), the $\Gamma$-distribution (Eq.~\ref{eq:app-eq}) approaches a Gaussian.  This is a direct consequence of the central 
limit theorem since the $\Gamma$-distribution is simply the reduced-$\chi^2$ distribution, i.e. $\chi^2/N_{\rm dof}$ for $N_{\rm dof}=2\alpha$ degrees of freedom, and 
hence represents the distribution of sample averages of $N_{\rm dof}$ squared unit Gaussian deviates.

\bibliography{mag100417}

\end{document}